\documentclass[3p,preprint]{elsarticle}

\usepackage{amssymb}
\usepackage{amsmath} % to use aligned environment

\usepackage{amssymb,amsthm,MnSymbol}
\usepackage{color}

\usepackage{algorithm}
\usepackage{algorithmic}

\usepackage{subfig}

\usepackage{amscd}
\begin{document}

\begin{frontmatter}



% \author[addr1]{}
\author[addr1]{I. Shevchenko\corref{cor1}}
\ead{i.shevchenko@imperial.ac.uk}
\author[addr1]{P. Berloff}
% \author[addr1,addr2]{P. Berloff}
 
%\fntext[label2]{}
\cortext[cor1]{Corresponding author at:}
%\address{Address\fnref{label3}}
%% \fntext[labecl3]{}

% \address{Department of Mathematics, Imperial College London, Huxley Building, 180 Queen's Gate, London, SW7 2AZ, UK.}
\address[addr1]{Department of Mathematics, Imperial College London, Huxley Building, 180 Queen's Gate, London, SW7 2AZ, UK}
% \address[addr2]{Institute of Numerical Mathematics of the Russian Academy of Sciences, Moscow, Russia}

% \dochead{Short communication}
%% e.g. \dochead{17th International Conference on Dynamical Processes in Excited States of Solids}

\title{A hyper-parameterization method for comprehensive ocean models: Advection of the image point}


%\author{}

%\address{}

\begin{abstract}
% \linenumbers
% \onehalfspacing % ???
Idealized and comprehensive ocean models at low resolutions cannot reproduce nominally-resolved flow structures
similar to those presented in the high-resolution solution. 
Although there are various underlying physical reasons for this, from the dynamical
system point of view all these reasons manifest themselves as a low-resolution trajectory avoiding the phase space
occupied by the reference solution (the high-resolution solution projected onto the coarse grid). 
In order to solve this problem, a set of hyper-parameterization methods has recently been proposed and successfully tested on idealized ocean models. 
% This study continues the series of works about the hyper-parameterization approach -- an alternative approach to the 
% eddy parameterization problem.
In this work, for the first time we apply one of hyper-parameterization methods (Advection of the image point) to 
a comprehensive, rather than idealized, general circulation model of the North Atlantic.

The results show that the hyper-parameterization method significantly improves a non-eddy-resolving solution towards the reference eddy-resolving solution 
by reproducing both the large- and small-scale features of the Gulf Stream flow. The proposed method is much faster than even a single run of the coarse-grid ocean model, 
requires no modification of the model, and is easy to implement. 
Moreover, the method can take not only the reference solution as input data
but also real measurements from different sources (drifters, weather stations, etc.), or combination of both.
All this offers a great flexibility to ocean modellers working with mathematical models and/or measurements.
\end{abstract}

\begin{keyword}
hyper-parameterization \sep build-up effect \sep parameterization \sep
comprehensive ocean model \sep primitive equations \sep large scale \sep small scale \sep eddy parameterization problem \sep nudging
\sep Gulf Stream flow



\end{keyword}

\end{frontmatter}

% \linenumbers

% \clearpage
% \tableofcontents


\section{Introduction}
The eddy parameterization problem (how to account for the effect of unresolved small scales onto the resolved large scales)
is one of the most challenging problems in the ocean modelling, counting decades of active research.
Despite a number of parameterizations (computationally affordable and physically justified mathematical models of 
unresolved processes) have been proposed  to solve the problem
(e.g.,~\citet{GentMcwilliams1990,DuanNadiga2007,Frederiksen_et_al2012,
PortaMana_Zanna2014,CooperZanna2015,
Grooms_et_al2015,Berloff_2015,Berloff_2016,Berloff_2018,Ryzhov_etal_2019,
CCHWS2019_1,Ryzhov_etal_2020,
CCHWS2019_3,CCHWS2020_4,CCHPS2020_J2}), it remains largely unresolved.
The main general point is that most of the parameterization approaches are physics-based rather than data-driven, and the former has obvious advantage of being valid in situation when the underlying physics changes.
However, in the situation when physics-based parameterizations remain in their infancy, there is a great practicality in considering data-driven parameterizations, 
which can reproduce nominally-resolved flow structures within their obvious limitations.
For example, for many research questions a computationally-cheap data-driven solution can replace a computationally-expensive dynamical ocean simulation in climate-type models and predictions.
Advancing the hyper-parameterization approach, within broader context of data-driven parameterizations, provides the main motivation for our study, in which we continue the series of precursor works~\citep{SB2021_J1,SB2022_J1,SB2022_J2}.
The main novelty is to adapt the methodology for fully comprehensive and realistic ocean models, and demonstrate its success with full confidence.
In turn this paves the way for broad use of hyper-parameterizations across the ocean modelling community.

% The main characteristic feature of many parameterization methods is a mathematical model describing the evolution of unresolved small scales and their effect
% on the resolved large scales. On the one hand, it is a strong side of physics-based parameterizations, as it does not require data (numerical simulations and/or measurements)
% to run the parameterized model, presumably, indefinitely long. Paradoxically, but it is their weakness as well (at least for now), since the small-scale machinery (i.e. how the 
% unresolved small scales work and how they effect the resolved large scales) still evades our understanding. 
% 
% This study continues the series of works~\citep{SB2021_J1,SB2022_J1,SB2022_J2} on the hyper-parameterization approach which aims to challenge 
% the eddy parameterization problem in the phase space, 
% as opposed to the mainstream parameterizations that work in the physical space.
% The main characteristic of the hyper-parameterization approach is that it does not require to know the physics of large-small
% scale interactions to model the flow dynamics at low resolutions and includes both data- and physics-driven methods. 

\section{The method}
% \section{The method and model configuration}
In this work, we consider the method called ``Advection of the image point'';
it is called so, because the image point is advected by the flow in phase space. 
The method falls into the category of data-driven methods, and has been successfully tested on
a multi-layer quasi-geostrophic model and showed promising results~\citep{SB2021_J1}. 
The method is based on the fact that the first-order ordinary differential equation
\begin{equation}
\mathbf{x}'(t)=\mathbf{F}(\mathbf{x}),\quad \mathbf{x}\in\mathbb{R}^n
\label{eq:ode1} 
\end{equation}
can be interpreted as a vector field $\mathbf{F}(\mathbf{x})$ in the phase space of equation~\eqref{eq:ode1}.
If $\mathbf{F}(\mathbf{x})$ is known, it can be used to advect the image point $\mathbf{x}$ in the phase space. Evolution of
an image point can be described by the equation:
\begin{equation}
\mathbf{y}'(t)=\frac{1}{N}\sum\limits_{i\in\mathcal{U}(\mathbf{y}(t))}\mathbf{F}(\mathbf{x}(t_i))+
\eta\left(\frac{1}{M}\sum\limits_{i\in\mathcal{U}(\mathbf{y}(t))}\mathbf{x}(t_i)-\mathbf{y}(t)\right),\quad
\mathbf{y}(t_0)=\mathbf{x}(t_0)\, ,
\label{eq:evolution_eq_no_nudging}
\end{equation}
where $\mathcal{U}(\mathbf{y}(t))$ is a neighbourhood of solution $\mathbf{y}(t)$,
and $i$ is the timestep of the reference solution $\mathbf{x}(t_i)$. The neighbourhood is computed as the average over 
$N$ (and $M$ for the nudging term) nearest, in $l_2$ norm, to the solution $\mathbf{y}(t)$ points, and $\eta$ is the nudging strength;
we will return to the choice of $N$, $M$, and $\eta$ in the next section; we refer to these parameters as hyper-parameters.
The hyper-parameters can be set based on the chosen metric 
and available data. Getting a bit ahead, we report that in our case the neighbourhood consists of $N=15$ and $M=5$ points, and the nudging strength is $\eta=0.001$.
Originally, the method was supposed to have the same size of the neighborhood for both the vector field $\mathbf{F}(\mathbf{x})$ and the nudging term (i.e., $N=M$). However,
our experiments showed that using neighborhoods of different sizes can prevent the so-called build-up effect which we discuss later.
% Our measure of goodness is how close the solution $\mathbf{y}(t)$ is to the reference phase space (the phase space occupied by the reference solution $\mathbf{x}(t)$). 
We refer the reader to~\citep{SB2021_J1} for a more detailed discussion of the method.

\section{Model configuration and numerical results}
\begin{figure}[H]
% \centering
\hspace*{-3.25cm}
\begin{tabular}{cccc}
& \hspace*{0.5cm}\begin{minipage}{0.33\textwidth} \hfill{$t=1$ year} \end{minipage} & 
\hspace*{-0.125cm}\begin{minipage}{0.33\textwidth} \hfill{$t=2$ years} \end{minipage} &
\hspace*{-0.125cm}\begin{minipage}{0.33\textwidth} \hfill{2-year average} \end{minipage}
\end{tabular}

\hspace*{2cm}
\begin{tabular}{cccc}
\hspace*{-5.75cm}\begin{minipage}{0.02\textwidth}\rotatebox{90}{\bf (a)}\end{minipage}  &
\hspace*{-3cm}\begin{minipage}{0.24\textwidth}\includegraphics[scale=0.45]{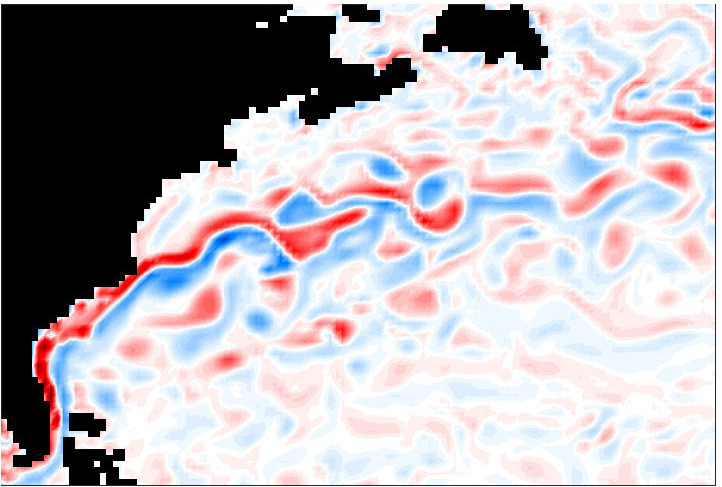}\end{minipage} &
\hspace*{1.25cm}\begin{minipage}{0.24\textwidth}\includegraphics[scale=0.45]{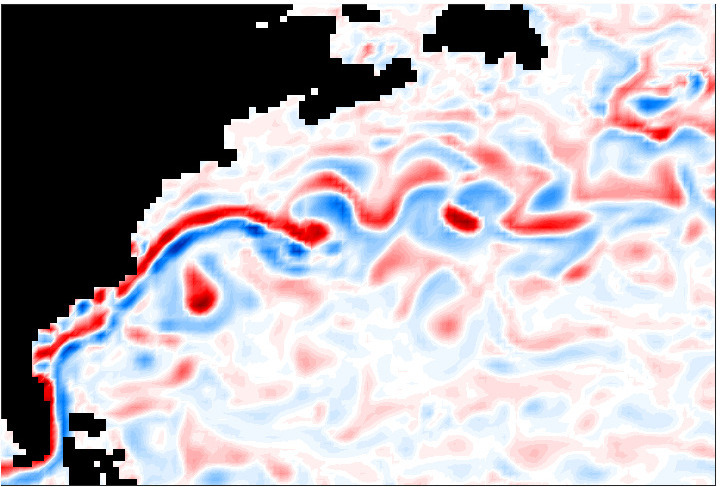}\end{minipage} &
\hspace*{1.3cm}\begin{minipage}{0.24\textwidth}\includegraphics[scale=0.45]{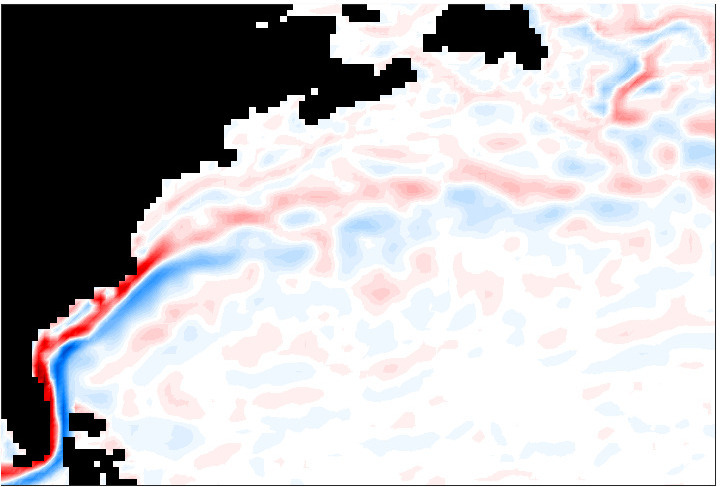}\end{minipage}\\
& & & \\[-0.35cm]
\hspace*{-5.75cm}\begin{minipage}{0.02\textwidth}\rotatebox{90}{\bf (b)}\end{minipage}  &
\hspace*{-3cm}\begin{minipage}{0.24\textwidth}\includegraphics[scale=0.45]{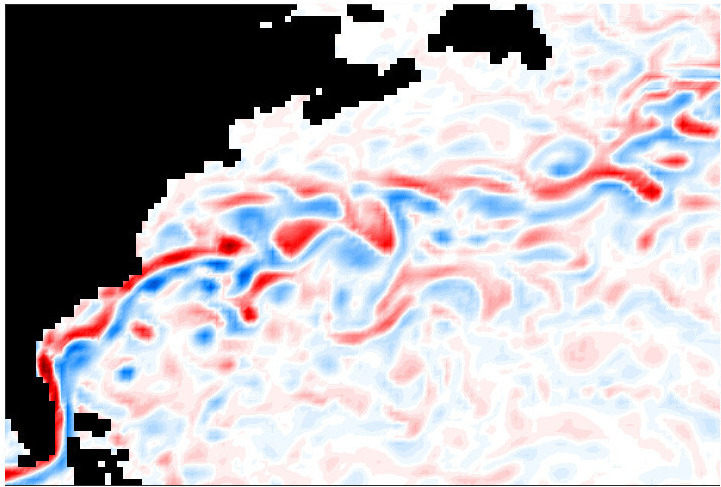}\end{minipage} &
\hspace*{1.25cm}\begin{minipage}{0.24\textwidth}\includegraphics[scale=0.45]{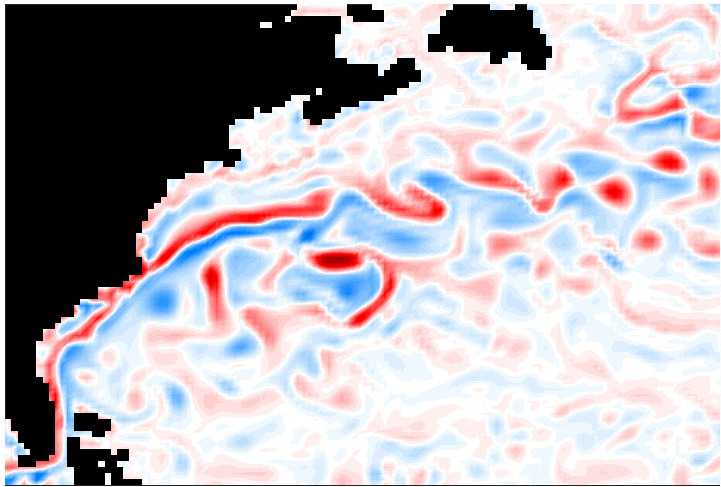}\end{minipage} &
\hspace*{1.3cm}\begin{minipage}{0.24\textwidth}\includegraphics[scale=0.45]{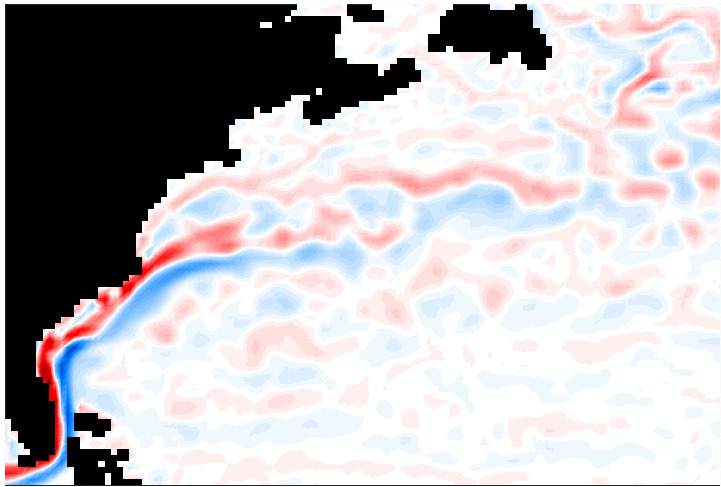}\end{minipage}\\
& & & \\[-0.35cm]
\hspace*{-5.75cm}\begin{minipage}{0.02\textwidth}\rotatebox{90}{\bf (c)}\end{minipage}  &
\hspace*{-3cm}\begin{minipage}{0.24\textwidth}\includegraphics[scale=0.45]{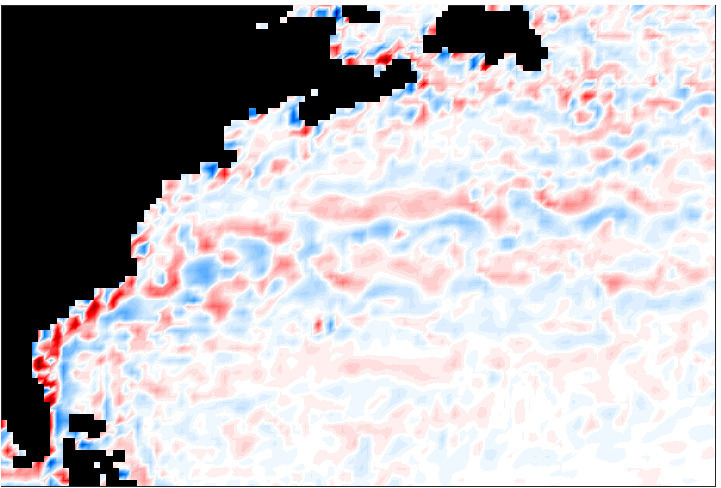}\end{minipage} &
\hspace*{1.25cm}\begin{minipage}{0.24\textwidth}\includegraphics[scale=0.45]{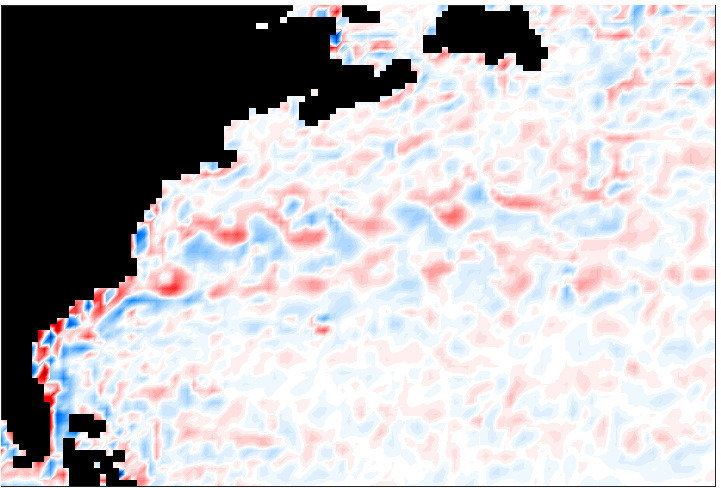}\end{minipage} &
\hspace*{1.3cm}\begin{minipage}{0.24\textwidth}\includegraphics[scale=0.45]{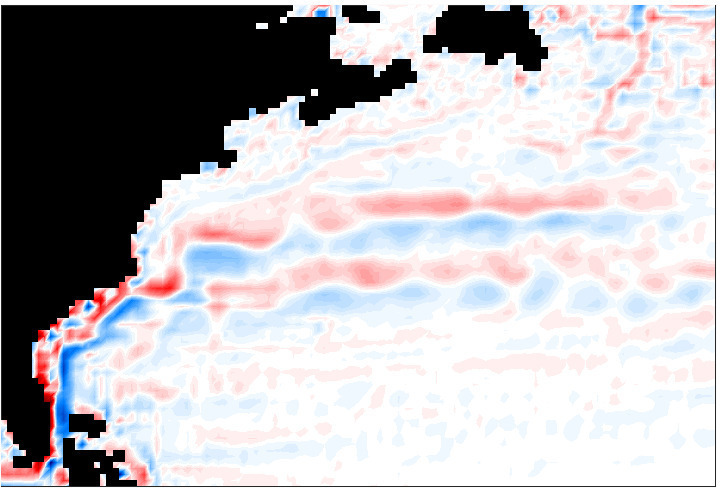}\end{minipage}\\
& & & \\[-0.35cm]
\multicolumn{4}{c}{\hspace*{-0.95cm}\includegraphics[width=6cm,height=0.75cm]{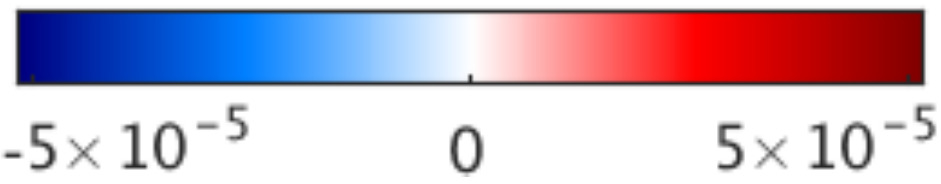}}\\
\end{tabular}
% 
% \vspace*{1cm}
% \hspace*{-3.25cm}
% \begin{tabular}{cccc}
% \multicolumn{4}{c}{\hspace*{2.5cm} \bf The build-up effect for the hyper-parameterized solution}\\
% & & & \\[-0.35cm]
% & \hspace*{0.5cm}\begin{minipage}{0.33\textwidth} \hfill{$t=1$ year} \end{minipage} & 
% \hspace*{-0.125cm}\begin{minipage}{0.33\textwidth} \hfill{$t=2$ years} \end{minipage} &
% \hspace*{-0.125cm}\begin{minipage}{0.43\textwidth} \hspace*{1.5cm} Time-average over the second year \end{minipage}
% \end{tabular}
% 
% \hspace*{2cm}
% \begin{tabular}{cccc}
% \hspace*{-5.75cm}\begin{minipage}{0.02\textwidth}\rotatebox{90}{\bf (d)}\end{minipage}  &
% \hspace*{-3.05cm}\begin{minipage}{0.24\textwidth}\includegraphics[scale=0.45]{RVbar_buildup1_1460.jpg}\end{minipage} &
% \hspace*{1.25cm}\begin{minipage}{0.24\textwidth}\includegraphics[scale=0.45]{RVbar_buildup1_2915.jpg}\end{minipage} &
% \hspace*{1.25cm}\begin{minipage}{0.24\textwidth}\includegraphics[scale=0.45]{RVbar_buildup1_time_av_second_half.jpg}\end{minipage}\\
% & & & \\[-0.35cm]
% \multicolumn{4}{c}{\hspace*{-0.95cm}\includegraphics[width=6cm,height=0.75cm]{colorbar_bwr.jpg}}\\
% \end{tabular}
\caption{Shown are snapshots of 
the surface relative vorticity $\zeta=v_x-u_y$ {\rm[1/s]} for {\bf (a)} the reference solution (computed 
at horizontal resolution $1/12^{\circ}$ and then projected on the $1/3^{\circ}$-grid), 
{\bf (b)} hyper-parameterized solution computed at horizontal resolution $1/3^{\circ}$ for $N=15$, $M=5$, $\eta=0.001$,
{\bf (c)} modelled solution computed at horizontal resolution $1/3^{\circ}$, % (started from the reference initial condition),
and the 2-year time-average (last column).
%, {\bf (d)} the build-up effect for the hyper-parameterized solution computed at horizontal resolution $1/3^{\circ}$ for $N=M=5$, $\eta=0.001$.
Snapshots are taken after 1 year (left column) and 2 years (middle column) of simulations.
Note the modelled solution {\bf (c)} fails to reproduce important large-scale 
(the Gulf Stream eastward jet extension) and small-scale (vortices) structures of the flow dynamics in both
instantaneous and time-averaged fields.}
\label{fig:rv_12_3}
\end{figure}

For the purpose of this study, we consider the Massachusetts Institute of Technology general circulation model (MITgcm)~\citep{Marshall_etal_1997} in the North Atlantic configuration~\citep{SB2022_J1}.
It is a 46-layer oceanic model coupled with an atmospheric boundary model~\citep{Marshall_etal_1997,Deremble_et_al2013}.
The coupled model is initially spun up for 5 years and then integrated for another 2 years. Although, 
for this work it is not essential that the initial state of 
the ocean circulation is in the statistically equilibrated regime.
The model is integrated at two different horizontal resolutions ($1/12^{\circ}$ and $1/3^{\circ}$);
the oceanic and atmospheric models are implemented with the same horizontal resolution.
We refer to the  $1/12^{\circ}$-solution projected onto the $1/3^{\circ}$-grid (Figure~\ref{fig:rv_12_3}a) 
as the reference solution, and to the solution computed on the $1/3^{\circ}$-grid as the modelled solution (Figure~\ref{fig:rv_12_3}c).
The hyper-parameterized solution computed with equation~\eqref{eq:evolution_eq_no_nudging} is presented in Figure~\ref{fig:rv_12_3}b. 
% Both the hyper-parameterized and the modelled solutions start from the reference initial condition. ???

As it follows from the results in Figure~\ref{fig:rv_12_3}, the hyper-parameterized solution computed on the coarse grid 
(Figure~\ref{fig:rv_12_3}b) reproduces both the large scales (the Gulf Stream flow) and small scales (vortices) of the reference solution 
(Figure~\ref{fig:rv_12_3}a) in 
instantaneous and time-averaged fields, while the solution computed on the coarse grid without the hyper-parameterization 
(Figure~\ref{fig:rv_12_3}c) leads to no Gulf Stream or small-scale flow features. 
We remark that the hyper-parameterized solution is a part of a 2-year simulation from which we use only
the first year of the reference solution; over the second year model~\eqref{eq:evolution_eq_no_nudging} runs on its own.
% Our results clearly demonstrate that the proposed hyper-parameterization method preserves large- (the Gulf Stream) and
% small-scale (vortices) flow patterns and shows no flow deterioration in contrast with the coarse-grid model.
% (caused by the build-up effect) when both the reference solution $\mathbf{x}(t)$ in the nudging term of~\eqref{eq:evolution_eq_no_nudging}  
% and the vector field $\mathbf{F}(\mathbf{x})$ in the right hand side of~\eqref{eq:evolution_eq_no_nudging} 
% are averaged over different number of the nearest (to the hyper-parameterized solution $\mathbf{y}(t)$) points.

\begin{figure}[H]
\hspace*{-2.5cm}
\begin{tabular}{ccc}
\multicolumn{3}{c}{\hspace*{2.5cm} \bf The build-up effect for the hyper-parameterized solution}\\
& & \\[-0.35cm]
\hspace*{0.5cm}\begin{minipage}{0.33\textwidth} \hfill{$t=1$ year} \end{minipage} & 
\hspace*{-0.125cm}\begin{minipage}{0.33\textwidth} \hfill{$t=2$ years} \end{minipage} &
\hspace*{-0.125cm}\begin{minipage}{0.43\textwidth} \hspace*{1.5cm} Time-average over the second year \end{minipage}
\end{tabular}

\hspace*{2.75cm}
\begin{tabular}{ccc}
\hspace*{-3.05cm}\begin{minipage}{0.24\textwidth}\includegraphics[scale=0.45]{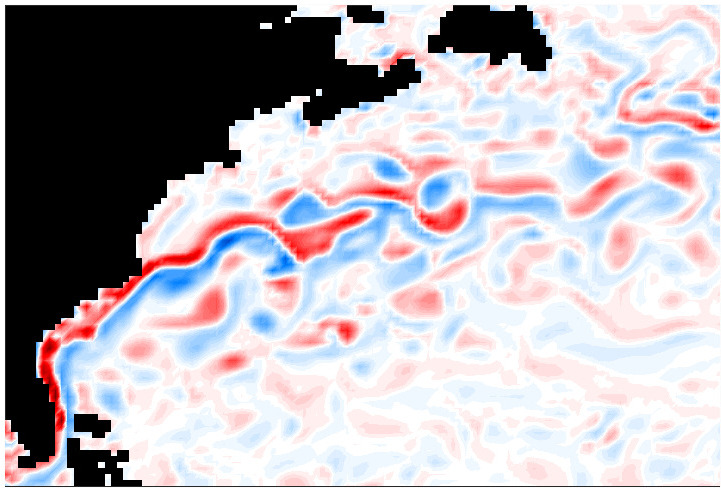}\end{minipage} &
\hspace*{1.25cm}\begin{minipage}{0.24\textwidth}\includegraphics[scale=0.45]{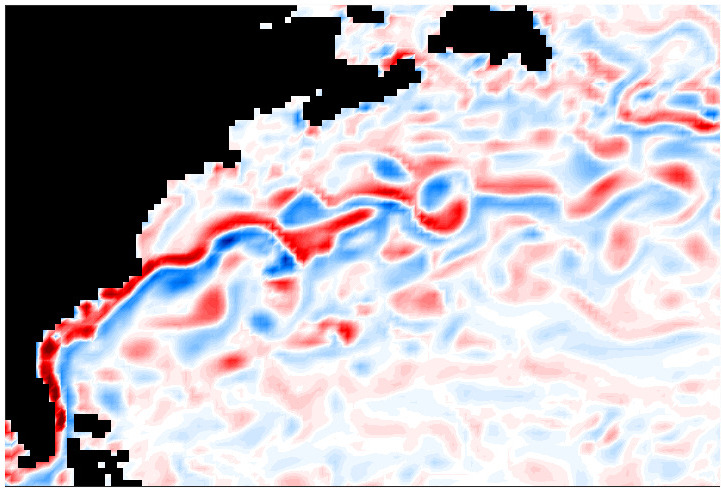}\end{minipage} &
\hspace*{1.25cm}\begin{minipage}{0.24\textwidth}\includegraphics[scale=0.45]{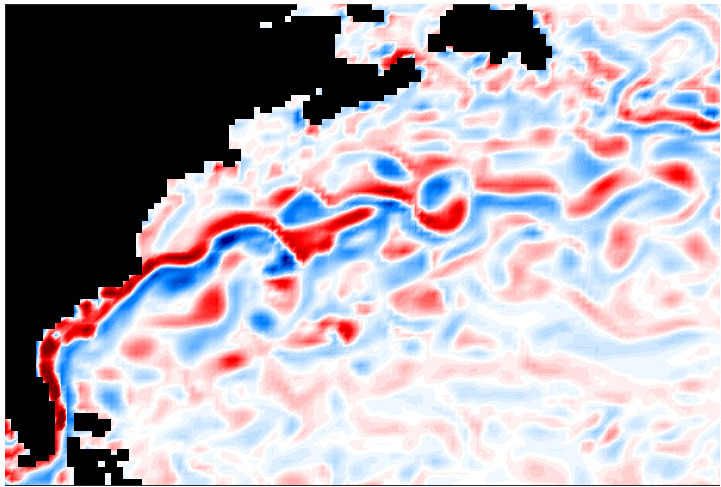}\end{minipage}\\
& & \\[-0.35cm]
\multicolumn{3}{c}{\hspace*{-1.35cm}\includegraphics[width=6cm,height=0.75cm]{colorbar_bwr.jpg}}\\
\end{tabular}
\caption{Shown are snapshots of the surface relative vorticity $\zeta=v_x-u_y$ {\rm[1/s]} demonstrating the build-up effect 
for the hyper-parameterized solution computed at horizontal resolution $1/3^{\circ}$  
for $N=M=5$, $\eta=0.001$.
Snapshots are taken after 1 year  (left column) and 2 years (middle column) of simulations.}
\label{fig:rv_12_3_buildup}
\end{figure}

The hyper-parameterized solution with a bad choice of hyper-parameters $N$, $M$, $\eta$ (Figure~\ref{fig:rv_12_3_buildup}) 
still reproduces both large- and small-scales flow features but only over the period in which the reference solution is available (it is one year in our case).
After that the solution almost stops evolving and eventually settles to a constant in time field like the one
in the middle plot of Figure~\ref{fig:rv_12_3_buildup}. It can be seen from 
the time-mean over the second year (the right subplot in Figure~\ref{fig:rv_12_3_buildup}), which is very similar to the snapshot 
taken at year 2 (middle subplot in Figure~\ref{fig:rv_12_3_buildup}), thus showing that the solution evolution is stalled over the second year.

{\bf The build-up effect and solution degradation}. 
The lack of reference data for the hyper-parameterized model and/or bad choice
of hyper-parameters can steer the trajectory to leave the reference phase space (the phase space occupied by the reference solution). 
This escape may result in a significant degradation of the hyper-parameterized solution and even lead to a numerical blow-up. We refer to this as the build-up effect, meaning 
that after a period of time, say $T$, the neighbourhood of the nearest points stalls (Figure~\ref{fig:rv_12_3_buildup}), i.e. the points in the neighbourhood become 
the closest ones to the image point $\mathbf{y}(t)$ for $\forall t>T$ (in the present case $T=1$ year); therefore, the same points are used again and again 
during the integration
of equation~\eqref{eq:evolution_eq_no_nudging} thus driving the image point away from the reference phase space. 
In principle, building up numerical errors may terminate this runaway, and the solution can return back to the reference phase space region, but this is case dependent and should be kept in mind. 
However, if the return time is relatively long (longer than the characteristic time of the reference solution) 
then the flow dynamics can be seriously distorted over the period of the trajectory injection.

In order to prevent the build-up effect, one should properly set up the hyper-parameters $N$, $M$, and $\eta$.
After some experiments we have found that $N=15$, $M=5$, and $\eta=0.001$ lead to no build-up effect (it is not necessarily the only choice, and other sets of hyper-parameters providing no build-up may exist).
As an alternative, one can also change the algorithm on how to pickup points from the neighbourhood. 
It is worth noting that these experiments are very fast and computationally cheap, even relative to a single run of the coarse-grid model.
We leave for the future work any further optimization of the image point advection algorithm.
% Over time and through experiments with a multi-layer quasi-geostrophic (QG) model
% and MITgcm we learned that the nudging methodology itself can be the remedy for the build-up. 
% For example, the build-up effect is observed in the hyper-parameterized QG solution which has been computed without nudging (Figure 5b in~\citep{SB2021_J1}). 
% However, when the nudging method is applied we observe no build-up effect (Figure 5c in~\citep{SB2021_J1}). 

\begin{figure}[h]
% \centering
\hspace*{-0.7cm}
\begin{tabular}{cc}
\hspace*{-0.5cm}\begin{minipage}{0.1\textwidth} {\bf (a)} \end{minipage} & \hspace*{-0.5cm}\begin{minipage}{0.1\textwidth} {\bf (b)} \end{minipage}\\
% \\[-0.35cm]
\begin{minipage}{0.5\textwidth}\includegraphics[scale=0.175]{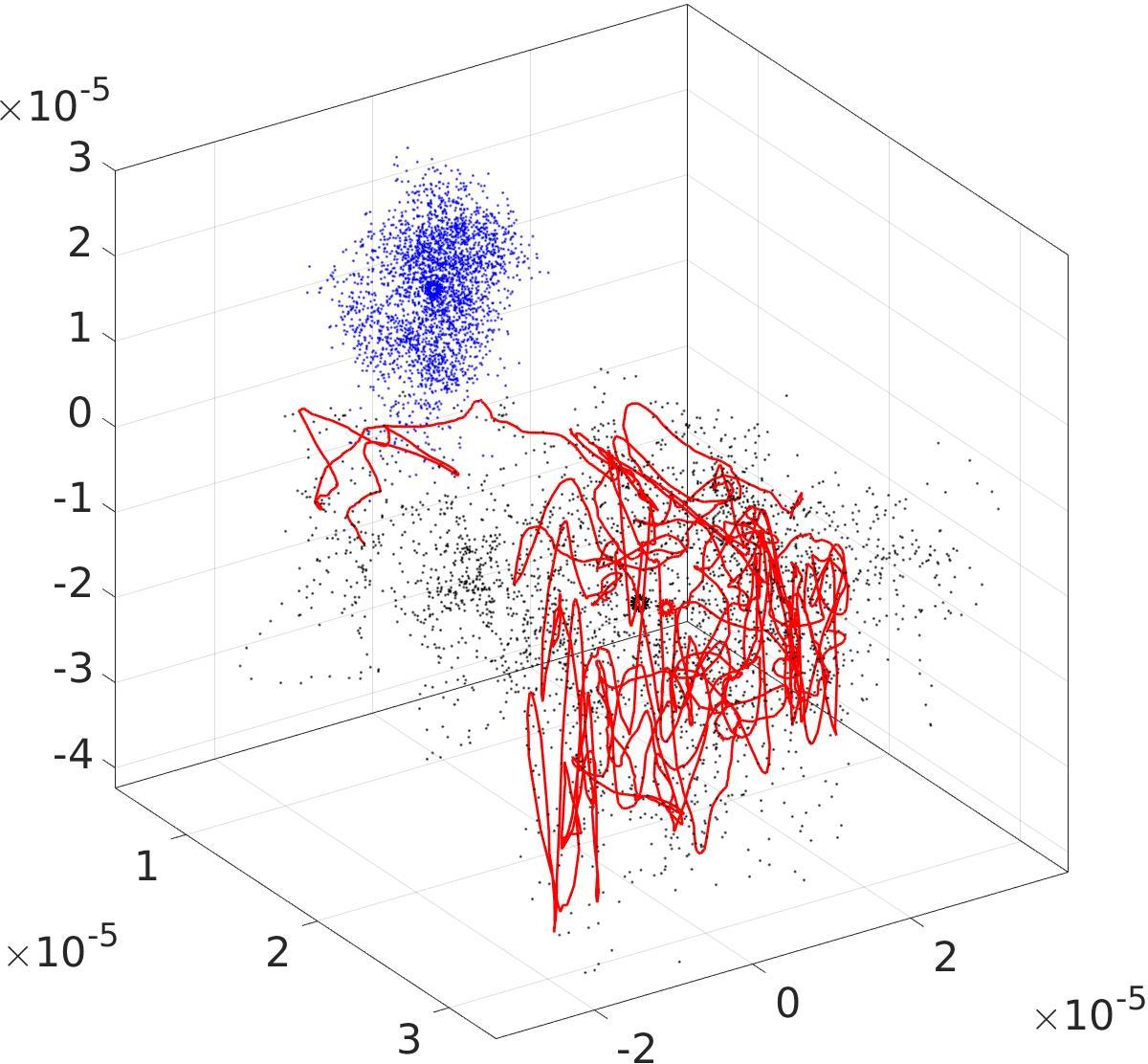}\end{minipage} &
\hspace*{-1cm}\begin{minipage}{0.5\textwidth}\includegraphics[scale=0.165]{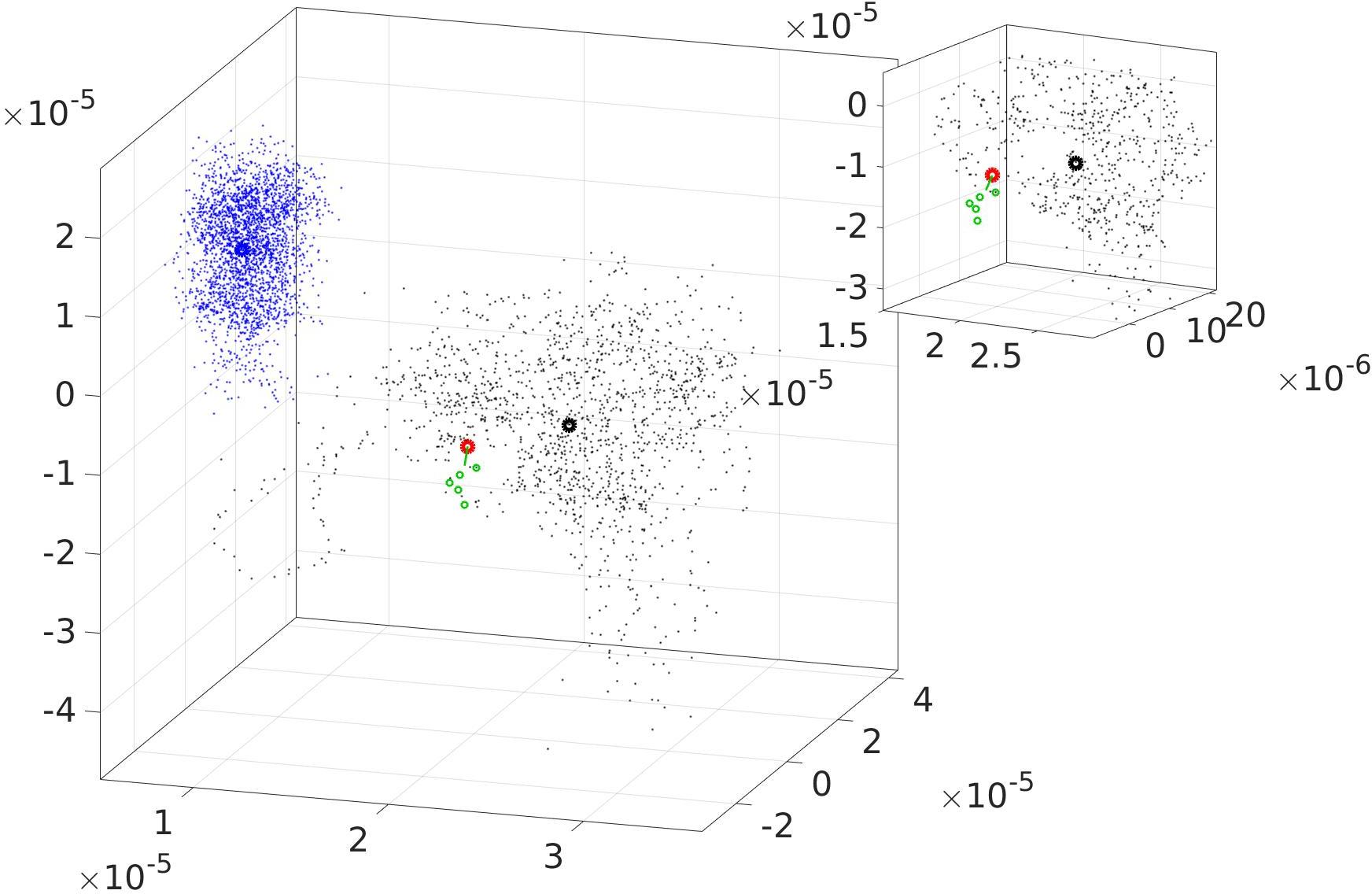}\end{minipage}\\
\end{tabular}
\caption{Shown is {\bf (a)} 3D projection of the reference phase space (black dots), modelled phase space (blue dots), hyper-parameterized
solution (red trajectory); in other words, we plot 3 random coordinates from the multi-dimensional phase space of surface relative vorticity,
{\bf (b)} build-up effect for the second year of the hyper-parameterized solution (short green line near the red circle)
for $N=M=5$, and $\eta=0.001$; the green circles are those stalled five points in the neighborhood.
The time-means for every solution are denoted by bold circles of corresponding colours.}
\label{fig:phase_space}
\end{figure}

{\bf The measure of goodness and evolution in phase space}.
The measure of goodness (i.e., the proximity of the modelled or parameterized solution to the reference one) in a given metric depends on the specific purpose.
Our measure of goodness is how close the hyper-parameterized solution % $\mathbf{y}(t)$ 
is to the reference phase space. %  (the phase space occupied by the reference solution $\mathbf{x}(t)$).
We use this measure to allow the hyper-parameterized solution to evolve in the neighborhood of the reference phase space, since
the failure of the coarse-grid model (blue dots in Figure~\ref{fig:phase_space}) to reproduce large- and small-scale features of the flow dynamics is because it steers away from where it should be
(black dots in Figure~\ref{fig:phase_space}, i.e. the reference phase space).
% As long as one ensures the proximity of the hyper-parameterized solution (red trajectory in Figure~\ref{fig:phase_space}) to the reference phase space (black dots in Figure~\ref{fig:phase_space}), it keeps evolving 
% in the right place of the phase space. 
Measuring the proximity of individual trajectories in phase space is of no value,
because small initial perturbations will grow exponentially due to the inherently chaotic dynamics.

\begin{figure}[h]
\centering  
\includegraphics[scale=0.175]{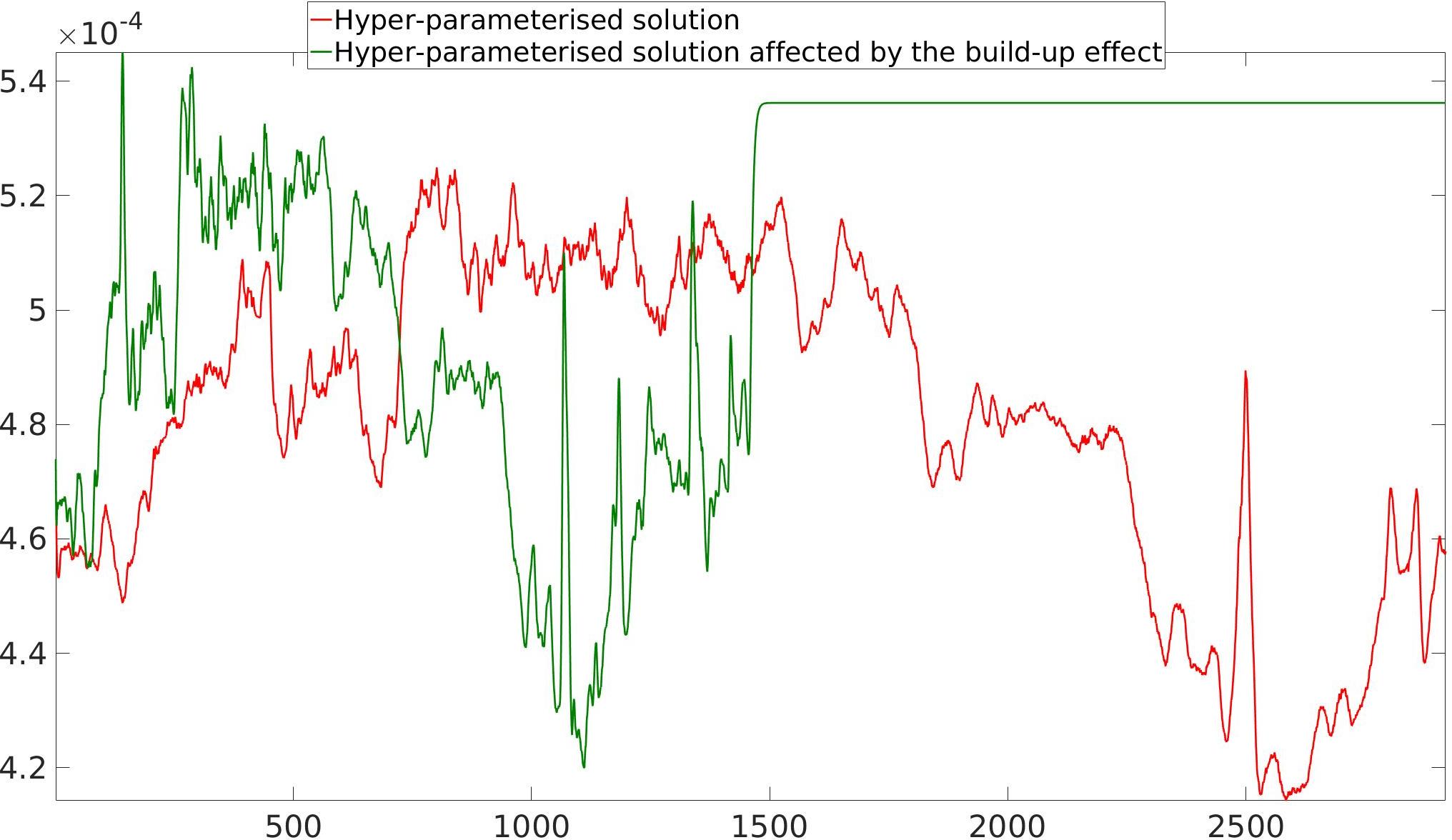}
\caption{Shown is evolution of the Euclidean distance (vertical axis) from the reference two-year time-mean 
to the hyper-parameterized solution (red line) 
and to the hyper-parameterized solution affected by the build-up effect (green line). 
The horizontal axis shows the simulation time in time steps (6 hours in our case).}
\label{fig:dist_to_time_mean}
\end{figure}

The build-up effect is demonstrated in Figure~\ref{fig:phase_space}b;
the short green line near the time-mean over the second year (red circle) is the 1-year evolution of the hyper-parameterized solution 
affected by the bad choice of hyper-parameters, namely $N=M=5$, and $\eta=0.001$. 
Evolution of the Euclidean distance from the reference time mean (the time-mean of the reference solution)
to the hyper-parameterized solution (Figure~\ref{fig:dist_to_time_mean}) shows that the hyper-parameterized solution 
affected by the build-up effect (green line) stops to evolve after one year (the period over which the reference solution is available).
In this case, we observe no numerical blow-up as the solution quickly settles to a constant in time field.
When the build-up effect is prevented, the hyper-parameterized solution (red line) continues to evolve with the reference phase space.

% What happens if only the nudging term is used? In this case, the solution drifts to the solution 
% $\frac{1}{M}\sum\nolimits_{i\in\mathcal{U}(\mathbf{y}(t))}\mathbf{x}(t_i)$.

\begin{figure}[H]
% \centering
\hspace*{-3.25cm}
\begin{tabular}{cccc}
\multicolumn{4}{c}{\hspace*{4.25cm} \bf Surface relative vorticity}\\
& \hspace*{0.5cm}\begin{minipage}{0.33\textwidth} \hfill{$t=1$ year} \end{minipage} & 
\hspace*{-0.125cm}\begin{minipage}{0.33\textwidth} \hfill{$t=2$ years} \end{minipage} &
\hspace*{-0.125cm}\begin{minipage}{0.33\textwidth} \hfill{2-year average} \end{minipage}
\end{tabular}

\hspace*{2cm}
\begin{tabular}{cccc}
\hspace*{-5.75cm}\begin{minipage}{0.02\textwidth}\rotatebox{90}{\bf (a)}\end{minipage}  &
\hspace*{-3cm}\begin{minipage}{0.24\textwidth}\includegraphics[scale=0.45]{RVref_11.jpg}\end{minipage} &
\hspace*{1.25cm}\begin{minipage}{0.24\textwidth}\includegraphics[scale=0.45]{RVref_21.jpg}\end{minipage} &
\hspace*{1.3cm}\begin{minipage}{0.24\textwidth}\includegraphics[scale=0.45]{RVref_31.jpg}\end{minipage}\\
& & & \\[-0.35cm]
\hspace*{-5.75cm}\begin{minipage}{0.02\textwidth}\rotatebox{90}{\bf (b)}\end{minipage}  &
\hspace*{-3cm}\begin{minipage}{0.24\textwidth}\includegraphics[scale=0.45]{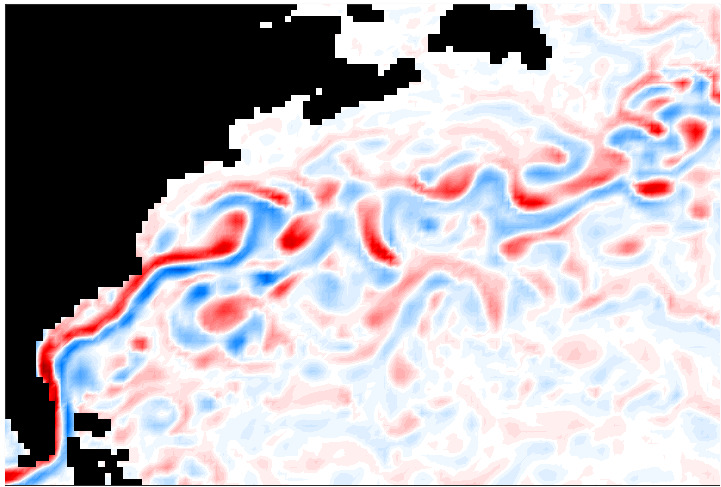}\end{minipage} &
\hspace*{1.25cm}\begin{minipage}{0.24\textwidth}\includegraphics[scale=0.45]{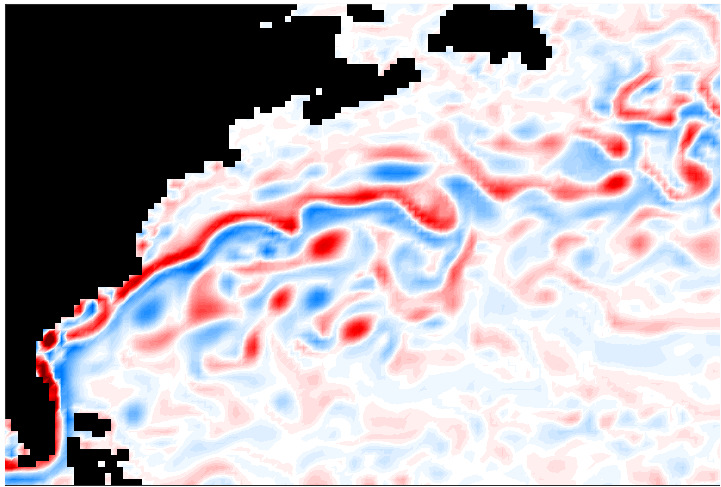}\end{minipage} &
\hspace*{1.3cm}\begin{minipage}{0.24\textwidth}\includegraphics[scale=0.45]{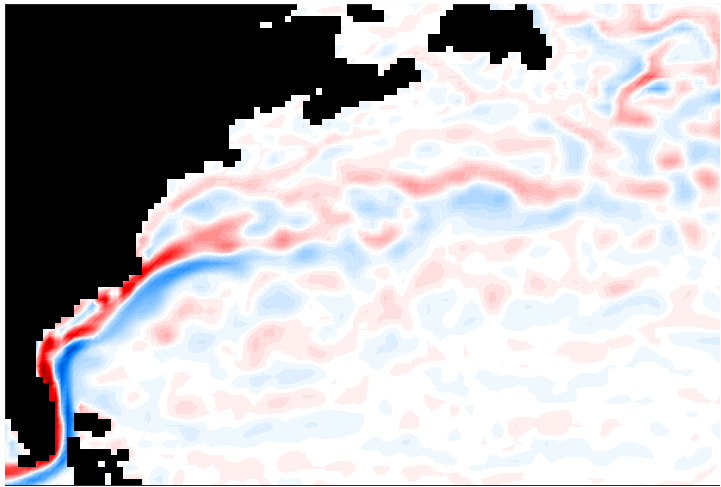}\end{minipage}\\
& & & \\[-0.35cm]
\multicolumn{4}{c}{\hspace*{-0.95cm}\includegraphics[width=6cm,height=0.75cm]{colorbar_bwr.jpg}}\\
\end{tabular}

\vspace*{0.25cm}
\hspace*{-3.25cm}
\begin{tabular}{cccc}
\multicolumn{4}{c}{\hspace*{4.25cm} \bf Sea surface temperature}\\
& \hspace*{0.5cm}\begin{minipage}{0.33\textwidth} \hfill{$t=1$ year} \end{minipage} & 
\hspace*{-0.125cm}\begin{minipage}{0.33\textwidth} \hfill{$t=2$ years} \end{minipage} &
\hspace*{-0.125cm}\begin{minipage}{0.33\textwidth} \hfill{2-year average} \end{minipage}
\end{tabular}

\hspace*{2cm}
\begin{tabular}{cccc}
\hspace*{-5.75cm}\begin{minipage}{0.02\textwidth}\rotatebox{90}{\bf (a)}\end{minipage}  &
\hspace*{-3cm}\begin{minipage}{0.24\textwidth}\includegraphics[scale=0.45]{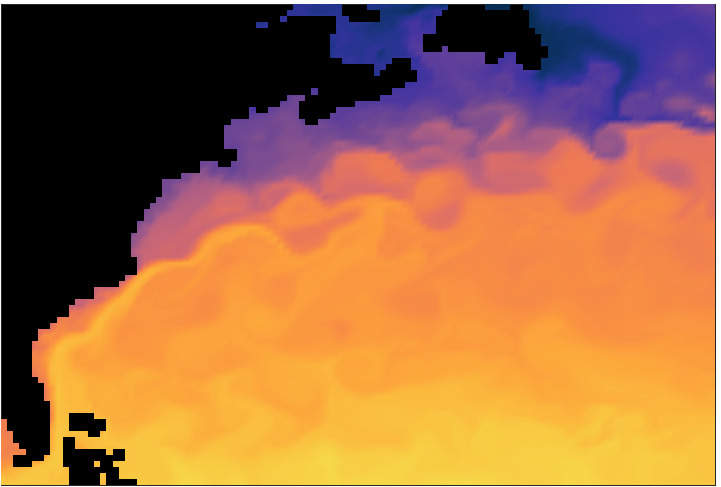}\end{minipage} &
\hspace*{1.25cm}\begin{minipage}{0.24\textwidth}\includegraphics[scale=0.45]{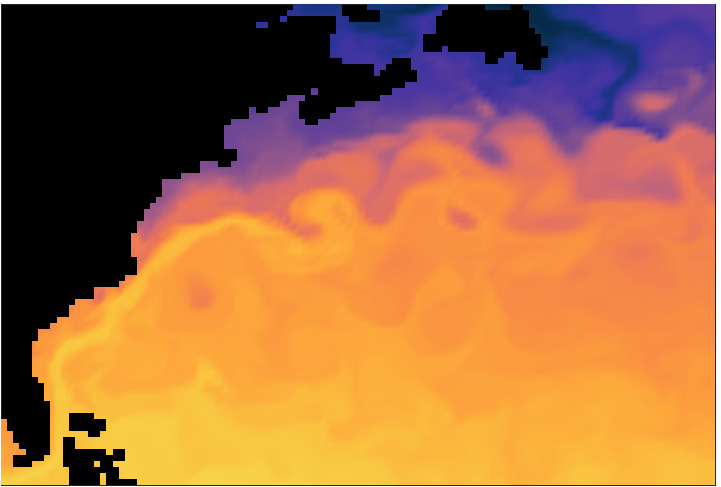}\end{minipage} &
\hspace*{1.3cm}\begin{minipage}{0.24\textwidth}\includegraphics[scale=0.45]{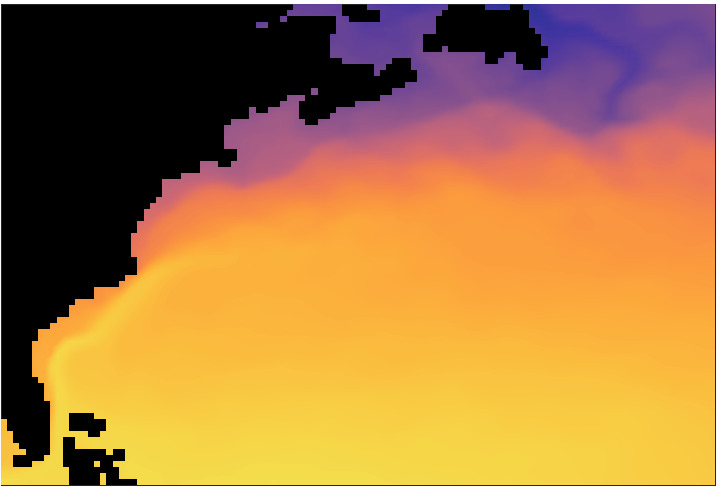}\end{minipage}\\
& & & \\[-0.35cm]
\hspace*{-5.75cm}\begin{minipage}{0.02\textwidth}\rotatebox{90}{\bf (b)}\end{minipage}  &
\hspace*{-3cm}\begin{minipage}{0.24\textwidth}\includegraphics[scale=0.45]{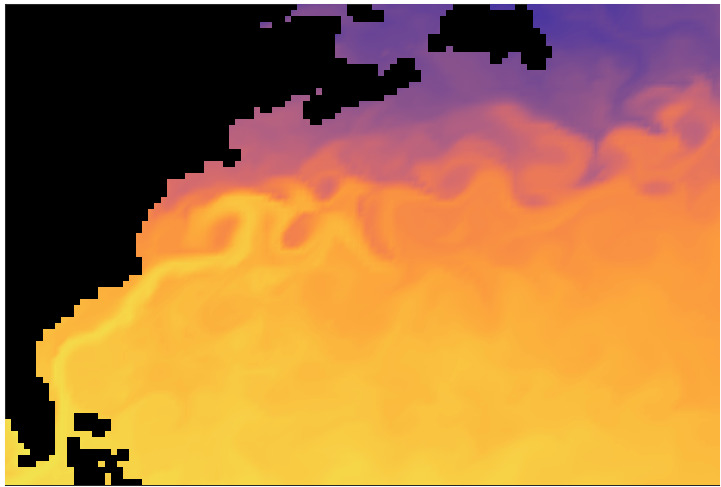}\end{minipage} &
\hspace*{1.25cm}\begin{minipage}{0.24\textwidth}\includegraphics[scale=0.45]{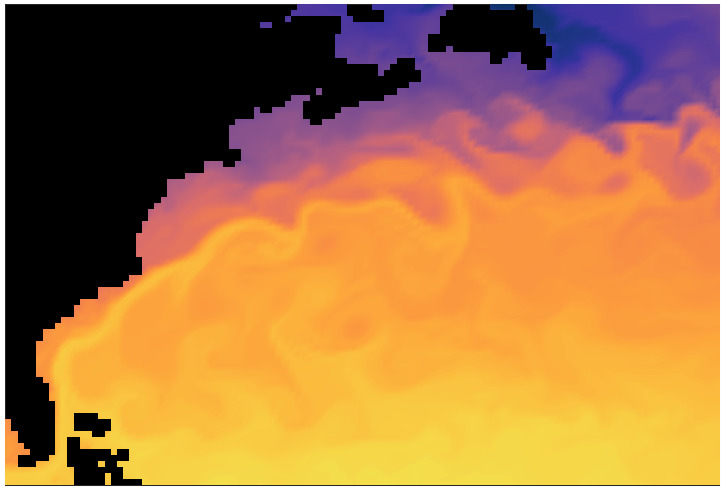}\end{minipage} &
\hspace*{1.3cm}\begin{minipage}{0.24\textwidth}\includegraphics[scale=0.45]{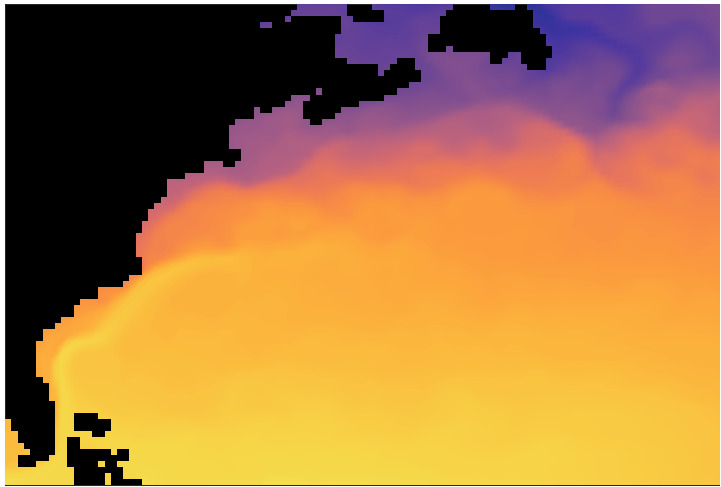}\end{minipage}\\
& & & \\[-0.35cm]
\multicolumn{4}{c}{\hspace*{-0.75cm}\includegraphics[width=6cm,height=0.75cm]{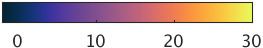}}\\
\end{tabular}
\caption{Shown are snapshots of the coupled fields
(sea surface relative vorticity $\zeta=v_x-u_y$ {\rm[1/s]} and surface temperature [$^\circ{\rm C}$])
for {\bf (a)} the reference solution (computed at horizontal resolution $1/12^{\circ}$ and then projected on the $1/3^{\circ}$-grid), 
{\bf (b)} hyper-parameterized solution computed at horizontal resolution $1/3^{\circ}$ for $N=18$, $M=4$, $\eta=0.001$,
and the 2-year time-average (last column).
Snapshots are taken after 1 year (left column) and 2 years (middle column) of simulations.
Note that the hyper-parameterized solution {\bf (b)} 
reproduces both large- (the Gulf Stream) and small-scale (vortices) features of the flow dynamics in both fields.}
\label{fig:rv_T_3}
\end{figure}

\clearpage
{\bf Coupled fields}. Comprehensive ocean models have several prognostic fields (velocity, temperature, etc.), and a good choice of
hyper-parameters ($N$, $M$, and $\eta$) for one field is not necessarily applicable to another.
For example, $N=15$, $M=5$, and $\eta=0.001$ choice works well only for the relative vorticity. However, it leads to a build up for the coupled fields (relative vorticity and temperature, not shown).
Thus, a set of hyper-parameters for a coupled case has to be found separately. 
Namely, for the coupled fields (relative vorticity and temperature) we have found that $N=18$ and $M=4$ remove
the build up effect, hence, the hyper-parameterized solution is of high quality (Figure~\ref{fig:rv_T_3}). 
The nudging strength in this case remains unchnged from the single relative vorticity case (Figure~\ref{fig:rv_12_3}), $\eta=0.001$.

This example of the coupled fields demonstrates that one should exercise caution when it comes to setting up the hyper-parameters.
On the other hand, it also shows that the hyper-parameterization method works for coupled fields (even with huge differences between the fields, which is 7 orders of magnitude for the relative
vorticity and temperature, see colorbars in Figure~\ref{fig:rv_T_3}).

\section{Conclusions and discussion\label{sec:conclusions}}

In this work we have applied the hyper-parameterization method ``Advection of the image point'' to 
the Massachusetts Institute of Technology general circulation model in the North Atlantic configuration and, thus, tested the method in a significantly
more complicated setting, as compared to the earlier idealized-model tests. Our results show that the hyper-parameterization 
method significantly improves a non-eddy-resolving solution (on a $1/3^{\circ}$-grid) towards the reference eddy-resolving solution 
(on a $1/12^{\circ}$-grid and then projected onto the $1/3^{\circ}$-grid) for both single and coupled fields (even with large difference between
the fields, it is 7 orders of magnitude in our case) by reproducing both the large-scale (the Gulf Stream eastward jet extension) and small-scale (vortices) flow features.
It is important to note that the hyper-parameterization method reproduces both large- and small-scale flow features not only over the period for which the reference solution is available
(1 year in our case), but also over the second year for which there is no reference data.

We have also explained the build-up effect and showed that a bad choice of hyper-parameters leads to the build-up effect,
and eventually to a significant degradation of the hyper-parameterized solution. 
In the worst-case scenario, the build-up can lead 
to a numerical blow up (which we did not observe in our experiments though). 
The build-up can be avoided by a proper setup of hyper-parameters which we have found through a series of numerical experiments.
In addition, the hyper-parameters have to be found separately
for hyper-parameterizing single and coupled fields, as well as for different single fields 
(for instance, the hyper-parameters that work well for relative vorticity may not be optimal for temperature, and vice versa).

It is important to keep in mind that the proposed method is data-driven and, therefore, can suffer from lack of data as any data-driven method. 
The hyper-parameterization approach has other methods~\citep{SB2022_J1,SB2022_J2} that
can be used to reproduce effects of mesoscale oceanic eddies on the large-scale ocean circulation, but demonstration of their skills on the level of comprehensive models is left for the future.
In other words, staying complimentary to the mainstream physics-based parameterization approach, we propose to work in phase space of the corresponding dynamical system and to interpret the lack of eddy effects as persistent tendency of phase space trajectories 
(representing the modelled solution) to escape the correct reference phase-space region.

The proposed method is much faster than even a single run of the coarse-grid ocean model, 
requires no modification of the model, and is easy to implement. The method can take as input data not only the reference solution
but also real measurements from different sources (drifters, weather stations, etc.), or combination of both.
All this offers a great flexibility not only to ocean modellers working with mathematical models but also to those working with measurements.

\section{Acknowledgments}
The authors thank The Leverhulme Trust for the support of this work through the grant RPG-2019-024. 
% and the anonymous referees for their constructive comments and suggestions,
% which have helped to improve the paper. 
% Pavel Berloff was supported by the NERC grants NE/R011567/1 and NE/T002220/1.
% , and by the Moscow Center for
% Fundamental and Applied Mathematics (supported by the Agreement 075-15-2019-1624 with the Ministry of Education
% and Science of the Russian Federation).



%% References


% \bibliographystyle{elsarticle-num}
\bibliographystyle{apalike}
\bibliography{refs}



\end{document}